\documentclass[11pt,a4paper]{article}
\usepackage{jcappub}
\usepackage{graphicx}
\usepackage{amssymb}
\usepackage{amsmath}
\usepackage{amsbsy}
\usepackage{color}
\usepackage{dsfont}
\usepackage{upgreek}
\usepackage{soul}
\usepackage[normalem]{ulem}
\newcommand{\ee}{\end{equation}}
\newcommand{\bb}{\begin{equation}}
\newcommand{\be}{\begin{equation}}
\newcommand{\eqb}{\begin{eqnarray}}
\newcommand{\eqf}{\end{eqnarray}}

\definecolor{gre}{rgb}{0,0.4,0.3}
\usepackage{chngpage}

\usepackage{color}

\title {\boldmath Vortex solutions of an Abelian Higgs model with visible and hidden sectors}

\author[a,b,1]{Paola Arias,\note{Corresponding author.}}
\author[c,2]{Fidel A.~Schaposnik\note{Also at CICBA.}}
\affiliation[a] {Instituto de F\'\i sica, Pontificia Universidad Cat\'olica de Chile, \\ Casilla 306, Santiago 22, Chile.}
\affiliation[b]{Departamento de F\'isica, Universidad de Santiago de Chile, \\Casilla 307, Santiago, Chile}
\affiliation[c]{Departamento de F\'\i sica, Universidad Nacional de La Plata,\\ Instituto de F\'\i sica La Plata, \\C.C. 67, 1900 La Plata, Argentina.}

\emailAdd{paola.arias@fis.puc.cl}
\emailAdd{fidelschaposnik@yahoo.co.uk}

\abstract{We study vortex solutions in a theory with dynamics governed by two weakly coupled Abelian Higgs models,   describing a hidden sector and a   visible sector. We analyze the radial dependence of the axially symmetric solutions constructed numerically and discuss the stability of vortex configurations for different values of the model  parameters, studying in detail vortex decay into lower energy configurations. We find that even in a weak coupling regime  vortex solutions  strongly depend on the  parameters of both the visible and hidden sectors. We also discuss on qualitative grounds possible implications of the existence of a hidden sector in connection with superconductivity and dark matter (dark strings).}
\begin{document}

\maketitle
\section{Introduction}

Models with  vector bosons and scalars in a hidden sector  naturally arise  in  supersymmetric extensions of the standard model as well as in superstring phenomenological studies.  They have also cosmological implications concerning gravitational wave production and dark matter abundance (see \cite{JR1} and references therein). Regarding this last issue, the hidden massive gauge boson could play the role of dark matter  \cite{Nelson:2011sf} or could be the messenger between the visible and dark sectors \cite{Arkani}. Also,
when the hidden sector has a $U(1)$ symmetry the corresponding gauge boson may have a very weak kinetic interaction with photons in the visible sector \cite{holdom}, which could lead  to observable effects in experiments like those on light shining through the wall, laser polarization and strong electromagnetic fields \cite{JR1}. Furthermore, when the hidden $U(1)$ gauge  symmetry is spontaneously broken the  classical field equations exhibit the well-honored Nielsen-Olesen vortex solutions that can play the role of dark strings   in an astrophysical context, as proposed in \cite{Vachaspati}-\cite{ Hyde}.

In view of the various areas in which the hidden sector could play an important role in explaining physical phenomena, it is of interest  to undertake the detailed study that we present in this work where we construct vortex solutions of two Abelian Higgs models associated to visible and hidden sectors weakly coupled through a gauge mixing interaction. In particular, we analyze how the effects of the hidden sector
depend not only on the strength of the mixing between the two $U(1)$ gauge bosons but also on the  relative strength of the gauge coupling constants and on   the scalar potentials parameters including the case   in which    one of the  $U(1)$ gauge symmetry remains unbroken. Another relevant subject that we analyze   concerns vortex decay. In the ordinary Abelian Higgs model vortex configurations with $n>1$ units of magnetic flux could decay  into elementary ($n=1$) vortices depending on the value of the Landau parameter \cite{JR}. We study this issue for configurations in which both hidden and visible vortices {{exist,}} and determine  how the mixing affects the decay scenario.

The plan of the paper is as follows: we introduce the model in section 2, extending the Nielsen-Olesen ansatz to include the hidden sector, leading to a coupled system of four radial field equations. In section 3 we consider the case in which the visible sector gauge symmetry is unbroken and discuss analytically  how the spontaneous breaking of the hidden sector gauge symmetry is communicated to the visible sector. Then, in section 4 we analyze numerically the case in which both the visible and hidden sectors gauge symmetries are broken studying the dependence of the vortex solutions  on the gauge mixing parameter (section 4.2) and on  the gauge coupling constants (section  4.3) using both a variational approach and a shooting method. Vortex decay is studied in section 5 and a  discussion of the   relevance of the model in connection with superconductivity is presented  in section 6. Section 7 gives a summary and discussion of our results.

\section{The model}
We consider a model with two $U(1)$ gauge fields, $A_\mu$ and $G_\mu$, each one coupled to  complex  scalars, $\phi$ and $\psi$ respectively, with dynamics governed by the following Lagrangian in $3+1$ space-time dimensions
\bb
\mathcal L=-\frac{1}{4} F_{\mu\nu}F^{\mu\nu}- \frac{1}{2}| D^\mu_A\phi|^2-V(\phi)-\frac{\chi}2F_{\mu\nu}G^{\mu\nu}-\frac{1}{4} G_{\mu\nu}G^{\mu\nu}- \frac{1}{2}| D^\mu_G\psi|^2 -V(\psi).
\label{1}
\ee
Here
\bb
\begin{array}{ll}
F_{\mu\nu} = \partial_\mu A_\nu - \partial_\nu A_\mu \;,  &  G_{\mu\nu} = \partial_\mu G_\nu - \partial_\nu G_\mu\\
~
\\
 D_{A}^\mu\phi=(\partial^\mu-ie A^\mu)\phi\;,   &  D_{G}^\mu\psi=(\partial^\mu-ie_h G^\mu) \psi
\end{array}
\ee
 and $V(\phi), V(\psi)$ are given by
\bb
V(\phi) =\frac{\lambda}4 \left(|\phi|^2-\phi_0^2\right)^2\, , \;\;\;\;\;\;
V(\psi)=\frac{\lambda_{\rm h}}4 \left(|\psi|^2-\psi_0^2\right)^2
\label{trez}
\ee
In our convention {{the set of}} fields  $A_\mu$ and $\phi$ belong to the visible sector,  while $G_\mu$ and $\psi$  belong to the hidden one. The strength of the mixing between the two gauge fields is parameterized by $\chi$ which could be either positive or negative.   Theoretical and  observational constraints seem to   favor at present that this parameter is small \cite{chismall,Essig:2013lka}. Although in principle $\chi$ is a free parameter, we show in the Appendix that consistency of boundary conditions  leading to the existence  of finite energy vortex solutions requires $|\chi|^2<1$.

We are interested in static configurations with $A_0 = G_0 = 0$  for which the energy density $\mathcal{E}$ associated to Lagrangian \eqref{1} takes the form
\bb
\mathcal{E}=\frac{B_i B_i}2+\frac{B_{hi}B_{hi}}2+\chi B_iB_{hi}+\frac{1}{2} |D^\mu_A\phi|^2+\frac{1}{2} |D^\mu_G\psi|^2+ V(\phi)+V(\psi) . \label{energy1}
\ee
with {{the magnetic fields of the visible and hidden sector defined as}}
\bb
B^i =  \varepsilon^{ijl}\partial_jA_k \;, \;\;\;\; B^{\,i}_{\!h}  = \varepsilon^{ijl}\partial_jG_k.
\ee
Due to the choice of symmetry breaking potentials, both gauge fields acquire masses {{given by}} $m_A^2=e^2 \phi_0^2$ and $m_G^2= e_h^2\psi_0^2$. Concerning the scalars, their masses are given by $m_\phi^2=2\lambda \phi_0^2$ and $m_\psi^2=2\lambda_h \psi_0^2$ according to the  Brout-Englert-Higgs mechanism.

It will be  convenient {{for later use}} to  define {{dimensionless}} coordinates, coupling constants and fields according to
\eqb
r\rightarrow r /  \left(e\phi_0\right), \,\,\, A_i\rightarrow \phi_0 A_i, \,\,\, \phi \rightarrow \phi_0\phi,\,\,\, G_i\rightarrow \phi_0 G_i,\,\,\, \psi \rightarrow \phi_0\psi.
\label{units}
\eqf
With this,  the energy per unit length $E/\ell$ in the $z$ direction, and with $A_z=G_z=0$, reads
\begin{eqnarray}
  \frac{E}{\ell} &=& \phi_0^2\int d^2 x\left\{\frac{B_iB_i}2+\frac{B_{hi}B_{hi}}2 +\frac{1}2 \left|\partial_i \phi -iA_i \phi\right|^2 +\frac{1}2 \left|\partial_i \psi -ie_rG_i \psi \right|^2\right.
\nonumber\\
&& + \left. \chi B_iB_{ hi} +V(|\phi|)+V(|\psi|)\right\} {\equiv \phi_0^2\int \!\!\! d^2 x \,{\tilde{\mathcal  E,}}}
\label{redefenergy}
\end{eqnarray}
where $e_r= e_h/ e$ and $\ell$ defines the length scale, $\ell = 1/e\phi_0$. The symmetry breaking potentials are now given by
\bb
V(|\phi|)=\frac{\kappa^2}8 \left(|\phi|^2-1\right)^2, \,\,\,\,\,\, V(|\psi|)=\frac{\beta^2 e_r^2}8 \left(|\psi|^2-\frac{\mu^2}{e_r^2}\right)^2.
\ee
{{Here we have defined a dimensionless parameter as }} $\kappa^2=2\lambda/ e^2$, which is related to  the  {\it Landau parameter} in the Ginzburg-Landau theory of superconductivity. {{The parameter}} $\beta^2=2\lambda_h/ e_h^2$, is its hidden analogue. Concerning {{the parameter}} $\mu$, it corresponds to the ratio of the hidden and visible gauge vector masses, $\mu =m_G/m_A= e_r \psi_0/\phi_0$.
In the ordinary Abelian Higgs model,  $\kappa <1$ corresponds to Type I superconductivity and $\kappa>1$  to Type II superconductivity. {{The limiting  value $\kappa=1$  is usually
called the Bogomolny point (for the ordinary Abelian Higgs model). At $\kappa=1$  can be derived}} a lower bound for the energy \cite{dVS, Bogo}. The bound is saturated  whenever the gauge and scalar fields  satisfy a system of coupled first order equations and the energy is then proportional to the number of quantized units of magnetic flux of the vortex solutions.

After {{the}} redefinitions {{stated in eq.~}}\eqref{units}
the visible gauge and scalar fields masses become  $m_A= 1$ and  $m_\phi= \kappa = \sqrt{{2\lambda}{/e^2}}$, respectively.  Concerning the hidden Higgs mass, it takes the {{value}}    $m_\psi= \sqrt{2\lambda_h \mu^2/e_h^2}$.
We are interested in finding static axially symmetric  solutions of the field equations,  so it will be convenient to consider
polar coordinates $(r,\varphi,z)$ and search for  $z$ independent field configurations. We then propose the well-honored Nielsen-Olesen \cite{NO}   ansatz  both for the visible and the hidden sectors
\bb
\begin{array}{ll|l}
\phi= \rho(r) e^{in\varphi},   & A_\varphi= n\frac{\alpha(r)}r, & A_r = 0,\,\,\,\,\, {{A_z=0,}}\,\,\,\,\,\,\,\,\,\,\, {{n\in \mathds{Z}.}} \\
~
\\
\psi= p(r) e^{ik\varphi},   &  G_\varphi= k\frac{\gamma(r)}{e_rr},  & G_r = 0,\,\,\,\,\, {{G_z=0,}}\,\,\,\,\,\,\,\,\,\,\, {{k\in \mathds{Z}.}}
\end{array}
\label{ansatz1}
\ee

Inserting  this ansatz, the energy density  (\ref{redefenergy})   in terms of  the redefined coordinates
 and parameters  \eqref{units} takes the form
\begin{eqnarray}
\tilde{\mathcal  E}\!\!&=&\!\!\frac{n^2}{2r^2}\left(\frac{d\alpha}{dr}\right)^2+\frac{k^2}{2e_r^2 r^2}\left(\frac{d \gamma}{dr}\right)^2+\chi \frac{nk}{e_r r^2}\frac{d \gamma}{dr} \frac{d \alpha}{dr}+\frac{1}2\left(\frac{d \rho}{dr}\right)^2+\frac{1}2\left(\frac{d p}{dr}\right)^2 \nonumber\\
&&\!\! {{+}}\frac{n^2}{2r^2} \left(1-\alpha\right)^2 \rho^2+\frac{k^2}{2r^2} \left(1-\gamma \right)^2 p^2+\frac{\kappa^2}8\left(\rho^2-1\right)^2+\frac{\beta^2 e_r^2}8\left(p^2-\frac{\mu^2}{e_r^2}\right)^2\!\!.\nonumber\\
\label{redefenergy2}
\end{eqnarray}
Finite energy density requires the following    behavior of fields at the origin and at infinity
\begin{eqnarray}
\rho(0) = p(0) = 0  \; , &&  \lim_{r \to \infty} \rho(r) = 1 \;, \;\;\; \lim_{r \to \infty}  p(r) = \frac{\mu}{e_r} \nonumber\\
 \alpha(0) = \gamma(0) = 0 \; , && \lim_{r \to \infty} \alpha(r)  =  \lim_{r \to \infty}  \gamma(r) = 1
 \label{boundary}
\end{eqnarray}
Using the asymptotic behavior   and  the fact that finite energy requires  covariant derivatives for both scalars to vanish at infinity one finds that the magnetic flux in the visible and hidden sectors can be written in terms of the scalar fields in the form
\begin{eqnarray}
\Phi_A &=& \oint_{{\cal C}_\infty} \!\!\!A_\mu dx^\mu  =  \frac{i}{e|\phi_0|^2} \oint_{{\cal C}_\infty} \!\!\! \phi^* \partial_\mu \phi\,  dx^\mu \,\,\,\,= \frac{2\pi n}e \,, \;\;\; n \in \mathbf{Z}
\label{primera}\\
\Phi_G &= &  \oint_{{\cal C}_\infty} \!\!\!G_\mu dx^\mu  = \frac{i}{e_h|\psi_0|^2} \oint_{{\cal C}_\infty} \!\!\! \psi^* \partial_\mu \psi\,  dx^\mu \!= \frac{2\pi k}{e_h}  \,, \;\;\; k \in \mathbf{Z}
\label{segunda}
\end{eqnarray}
 Here  the fluxes are written in terms of the original fields introduced in eqs.\eqref{1}-\eqref{trez}, i.e.~before redefining coordinates, coupling constants and fields.

Given ansatz \eqref{ansatz1}, the field equations for the model take the form
\eqb
&& \hphantom{-}n \alpha''+\chi\frac{k}{e_r} \gamma'' -\chi\frac{k}{e_r} \frac{\gamma'}r-n\frac{\alpha'}r  -n\left(\alpha-1\right)\rho^2 =0.  \label{diffeq1} \\
 && \hphantom{-} \frac{k}{e_r} \gamma''+n\chi \alpha''-\frac{k}{e_r}\gamma'-\chi n\frac{\alpha'}r-ek\left(\gamma-1\right)p^2=0. \label{diffeq2} \\
 &&\hphantom{-}\frac{1}r \frac{d}{dr}\left(r \rho'\right)-{{\frac{n^2}{r^2}\left(1-\alpha\right)^2}} \rho-\frac{\kappa^2}2\left(\rho^2-1\right)\rho=0.\label{order field} \label{diffeq3}\\
 &&\hphantom{-}\frac{1}r \frac{d}{dr}\left(r p'\right)-{{\frac{k^2}{r^2}\left(1-\gamma\right)^2}} p-\frac{\beta^2 e_r^2}2\left(p^2-\frac{\mu^2}{e_r^2}\right)p=0.\label{disorder field}
\eqf
where the prime indicates from now on a derivative with respect to $r$.

{Equations (\ref{diffeq1})-(\ref{disorder field}) decouple in the asymptotic regime where analytic solutions can be easily found. The asymptotic solution for $\alpha(r)$ and $\gamma(r)$ is encoded in the equation
\bb
 \left[ r\frac{d}{dr}  \left(\frac{1}r \frac{d}{dr}\right) \right] F_{\pm }=\frac{1}{\sqrt{C_{\pm}}} F_{\pm },
 \ee
where $F_\pm$ are a linear combination of $\alpha$ and $\gamma$ {{and $C_\pm$ are  coefficients depending on  $\chi$ and $\mu$}} (see appendix for details). {{Finite energy per unit length solutions require}}
$
\chi^2 <1
$.
Thus, in order to have  finite energy vortex solutions{{, the} parameter $\chi$ - controlling the mixing between the visible and hidden sectors - should satisfy $|\chi|<1$.}

Due to the presence of the gauge kinetic mixing no first-order Bogomolny equations \cite{dVS, Bogo} can be found when $\chi \ne 0$, except for a very particular case \cite{betti}.  {Evidently,  if    fields and parameters in the visible and the hidden sector are identified (this implying also the the number of units of magnetic flux in the ansatz)}, Lagrangian  \eqref{redefenergy} becomes the same as that of the ordinary
Abelian Higgs model apart from an overall factor $1/2$ and a shift in the gauge coupling constant $e \to e/\sqrt{1 - \chi^2}$. Hence, in this very special case one finds the usual Bogomolny equations with the Bogomolny point  separating Type I and Type II superconductivities shifted accordingly, $\kappa ^2 \to  (1 - \chi^2)\kappa^2$. We shall not discuss this case in which visible and hidden sectors become indistinguishable since it escapes the main interest of our work.

\section{One unbroken $U(1)$ symmetry}

Let us start by studying the existence and stability of vortex solutions  when one of the $U(1)$ gauge groups remains unbroken.
A related discussion has been presented in \cite{betti}, but we include the analysis here for completeness and to
highlight certain features that the model exhibits and we consider of interest.

Let us assume that the visible $U(1)$ gauge group remains unbroken (we could have chosen the other way around as well).
The simplest way to achieve this {is by eliminating the  visible scalar sector so that all $\phi$ dependent terms  in Lagrangian \eqref{1} are absent}.

 The energy density then reads
\bb
{\mathcal  E}_{U(1)} =  \frac{B_iB_i}2+\frac{B_{ hi}B_{hi}}2+ \chi B_iB_{hi}  +\frac{1}2 \left|\partial_i \psi -ie_h G_i \psi\right|^2
 +V(|\psi|)
 \label{originalE}
\ee
 We {{now perform a redefinition of the visible magnetic field as}} 
 \bb
 B_i = \tilde B_i - \chi B_{hi},
 \label{Bredef}
 \ee
 the energy density ${\mathcal  E}_{U(1)} $  becomes
 \bb
{\mathcal  E}_{U(1)} = (1  - \chi^2) \frac{B_{hi}B_{hi}}2+\frac{\tilde B_i\tilde B_i}2   +\frac{1}2 \left|\partial_i \psi -ie_h G_i \psi\right|^2
 +V(|\psi|).
 \label{enB}
\ee

Now, a redefinition of the hidden {{vector}} field, as
\be
G_i = \frac{G'_i}{\sqrt{1- \chi^2}}, 
\ee
{{leads to $B_{hi}={{B'_{hi}}}/\sqrt{1-\chi^2}$}}. We can rewrite  the energy (\ref{enB}) {{in terms of the new fields as}} 
\bb
{\mathcal  E}_{U(1)} =  \frac{{B'_{{h}i}}B'_{{h}i}}2+\frac{\tilde B_i \tilde B_i}2  +\frac{1}2 \left|\partial_i \psi -ie_{\rm eff}G'_i \psi\right|^2
 +V(|\psi|),
 \label{enB2}
\ee
where {{we have defined an}} effective coupling constant  $e_{\rm eff}$,  for the hidden gauge field 
 \be
 e_{\rm eff} = \frac{e_h}{\sqrt{1 - \chi^2}}.
 \label{bis}
 \ee

 Let us note that {{in terms of the redefined fields,}}  the energy density is the sum of two uncoupled terms: the one corresponding to the hidden sector coincides with the ordinary  Nielsen-Olesen vortex energy density, while the other one is just a Maxwell term for the  $\tilde B$ magnetic field. In this form, the energy density can be written as a sum of squares {whenever coupling constants are accommodated to fulfill  the Bogomolny condition}
\eqb
\begin{aligned}
 E/\ell=\psi_0^2\int d^2 x\frac{1}4\left\{\left(G'_{ij}\pm \varepsilon_{ij}\left(\psi^a\psi^a-1\right) \right)^2 + \left(D_i\psi^a \mp \varepsilon^{ab}\varepsilon_{ij}D_j\psi^b\right)^2\right.\\
+ \left. 4\left(\frac{\beta_h}2-\frac{1}2\right)\left(\psi^a\psi^a-1\right)^2\pm \left( \varepsilon_{ij}G'_{ij}\mp \varepsilon^{ab}\varepsilon_{ij}\partial_i\left(\psi^aD_j\psi^b\right)\right)
+ \tilde B_i \tilde B^i \right\}.
\end{aligned}
\eqf
Where we have moved to the dimensionless variables, $r\rightarrow r/(e_{{{\rm{eff}}}} \psi_0)$, $G'_i\rightarrow G'_i \psi_0$, $\psi \rightarrow \psi\psi_0$, $\tilde A_i\rightarrow \psi_0\tilde A_i$.

The minimization of the energy is bounded from below to
\bb
E/\ell\geq \psi_0^2\frac{2\pi}{e_{\rm eff}}k, \,\,\,\,\,\,\,\,\,\,\,\,\, k\in \mathds{Z}.
\ee
The bound is saturated when the  following set of Bogomolny equations are satisfied
\eqb
G'_{ij}&=&\mp \varepsilon_{ij}\left(\psi^a\psi^a-1\right).\\
D_i\psi^a&=&\pm \varepsilon_{ij}^{ab} D_j\psi_b.\\
\frac{1}2\varepsilon_{ij}\tilde F_{ij}&=&0.
\eqf
Thus, the configuration of minimum energy is the one where $\tilde B=0$. Going back to the original field of eq.~(\ref{Bredef}),
\bb
B=-\chi B_h.
\label{identity}
\ee
This result shows that even in the absence of symmetry breaking, the mixing between the visible and the hidden gauge field forces the former
to form a vortex with the same winding number $k$ as  the broken gauge field  hence having a quantized magnetic flux
\be
\Phi  = \oint_{{\cal C}_\infty} \!\!\!A_\mu dx^\mu =-\frac{\chi}{e_{{{\rm{eff}}}} }  2\pi k.
\ee
{Relation \eqref{identity} between both gauge fields implies that even in the absence of a symmetry breaking of the visible sector, the kinetic gauge mixing forces the magnetic field to have an exponential decay controlled by the hidden gauge field mass. Now, since in this case  the visible magnetic field $B$ is related to the hidden one according to $B = -\chi B_h$,  its strength is diminished by the kinetic mixing parameter.}

 This result could have   interesting phenomenological implications if this model is considered as providing a mixing of hidden and visible cosmic strings in the early universe.\footnote{It has been noted that cosmic strings produced during  phase transitions could seed primordial magnetic fields \cite{Vachaspati:1991nm}. One could think in a physical scenario where dark strings are formed during phase transition of the hidden sector, and as a consequence of the mixing, visible cosmic strings are formed, which in turn could seed a primordial magnetic field.}

Note that a similar topological effect for the dark and visible magnetic charge relation can  take place, as described in \cite{Brummer:2009cs}.

{When $B$ is an external field,   $\tilde B=0$ is no {{longer}} a solution, and  the role of the kinetic mixing is to lower the magnetic energy of the visible sector, as noted in \cite{betti}}.

\section{Numerical results}
{We shall first solve equations \eqref{diffeq1}-\eqref{disorder field} using a simple and effective variational approach that has been shown to render the energy of vortex solutions with similar accuracy as more elaborated methods \cite{hill}. Using this approach we shall analyze the dependence of the energy on the kinetic mixing parameter $\chi$ and the gauge coupling constants.
We shall also solve the field equations   using an asymptotic shooting method in order to obtain accurate profiles of the gauge and scalar field  vortex configurations.}
\subsection{Variational analysis}
The idea is to combine  powers of exponentials   to engineer  functions $\rho, \alpha, p$ and $\gamma$
with the  short- and long-distance limits imposed by conditions \eqref{ansatz1}
\bb
\begin{array}{ll}
\alpha(r)= \left(1-e^{-u r}\right)^2, & \rho(r)= 1-e^{-hr} \\
\gamma(r)=\left(1-e^{-f r}\right)^2, &p(r)=\frac{\mu}{e_r} \left(1-e^{-v r}\right). \label{ansatfi}
\end{array}
\ee
Variational parameters  $u$ and $f$ are related to the   visible and hidden gauge field masses respectively while $h$ and $v$ are related to the masses of the visible and hidden Higgs fields. In terms of these variational parameters $\tilde{\mathcal  E}$ takes the form:
\eqb
\tilde{\mathcal  E}\!\!\!\!&=&\!\!\!\!\frac{k^2}{2e_r^2}\left(\frac{e^{-4fr}}{r^2}\left(\mu ^2 \left(1-2 e^{f r}\right)^2
   \left(1-e^{-r v}\right)^{2 |k|}+4 f^2
   \left(e^{f r}-1\right)^2\right)\right.\nonumber\\
  &&\!\!\!\!\left. +\,{4 v^2\mu^2}e^{-2vr} \left(1-e^{-rv}\right)^{2|k|-2}\vphantom{\frac{e^{-4fr}}{r^2}}\right)+
  \frac{n^2}2\left(\frac{e^{-4ur}}{r^2}\left( \left(1-2 e^{u r}\right)^2
   \left(1-e^{-hr}\right)^{2 |n|}\right.\right.\nonumber \\ \nonumber
  &&\!\!\!\!\left. \left.
   +4 u^2
   \left(e^{u r}-1\right)^2\right)+{4 h^2}e^{-2hr} \left(1-e^{-hr}\right)^{2|k|-2}\vphantom{\frac{e^{-4fr}}{r^2}}\right)\\ \nonumber
   &&\!\!\!\!+ nk \frac{4uf\chi }{e_r r^2}\left(e^{rv}-1\right)\left(e^{fr}-1\right) e^{-2r(f+u)}+ \frac{\beta^2 }{8} \frac{\mu^4}{e_r^2} \left( \left(1-e^{-rv}\right)^{2 |k|}-1\right)^2\nonumber\\
   &&\!\!\!\!+\frac{\kappa^2 }{8}  \left(\left(1-e^{-h r}\right)^{2
   |n|}-1\right)^2
   \label{energyvar}
\eqf

Apart from the variational parameters, there are seven free parameters which should be chosen on physical grounds: $\kappa$ and  $\beta$, related to the Landau parameters for both the visible and hidden sector, $e_r = e_h/e$ related  to gauge coupling constants,  $\mu = m_G/m_A$, the ratio of gauge field masses, $\chi$ which measures $A_\mu$ and $G_\mu$ mixing strength   and $n$, $k$,  the number of units of visible and hidden magnetic fluxes.

To start with and in order to test our variational approach, we have considered the case in which there is no mixing ($\chi = 0$) for which we
have direct comparison with
 very accurate numerical results \cite{JR,dVS}}. We found that there is excellent
agreement between {those results  and ours. As an example, exact $n=1$ vortex energy
per unit length at the Bogomolny point is  $E/{\ell}{{=|\phi_0|^2}}$, while that obtained in ref.~\cite{JR} using a refined {{variational method}} is $E/{\ell} = 1.00000 |\phi_0|^2$. Concerning our  simpler variational result,  we obtained $E/{\ell} = {{1.01823}} |\phi_0|^2$.  In short, we trust the results of our variational calculation.}

{When the mixing between hidden and visible vector fields is so small that it can be ignored, the visible and hidden terms in the
model defined by Lagrangian (\ref{1})
decouple, and then there  exist two unrelated vortex solutions with winding numbers $n,k$, respectively. Recall that  when there is just one gauge field and one complex scalar and the Landau parameter is larger than the value   it takes at the  Bogomolny point, a vortex with winding number   $n>1$ decays into separated vortices
(see for example \cite{JR}) and this is then true for each of the  decoupled sector we refer above.}

{As we shall see, non negligible values of the kinetic mixing parameter $\chi$ can have  great impact in the existence of  vortex solutions and  their behavior.
 This will be also  the case concerning different values of the
 hidden gauge coupling constant  $e_r$,  {and/or}  the hidden gauge field mass appearing in the $\mu$ parameter}.

\subsection{Changing $\chi$}

Here we study  vortex stability as a function of {{the mixing parameter}} $\chi$.  As highlighted in the previous section, when the mixing parameter vanishes we are left with two uncoupled vortices, if their  Landau parameters  $\lambda, \beta $  are greater than one  {{and the corresponding winding numbers are greater than one, }}they become unstable, {{decaying in configurations of smaller winding numbers}}.

We shall show here that if the mixing is non negligible,  the stability conditions change and instability could take place without requiring    that both ($\lambda, \beta) >1$ simultaneously.
 To see this we fix  $\kappa$ to the value  it takes at  the stability critical point (Bogomolny point \cite{dVS, Bogo}) in the absence of mixing,  when  the theory  reduces to two uncoupled  Abelian Higgs model  exhibiting independent vortex solutions.  We then study the energy as a function of the hidden Landau parameter $\beta$ for different values of $\chi$.

\begin{figure}
\centering
\includegraphics[scale=0.7]{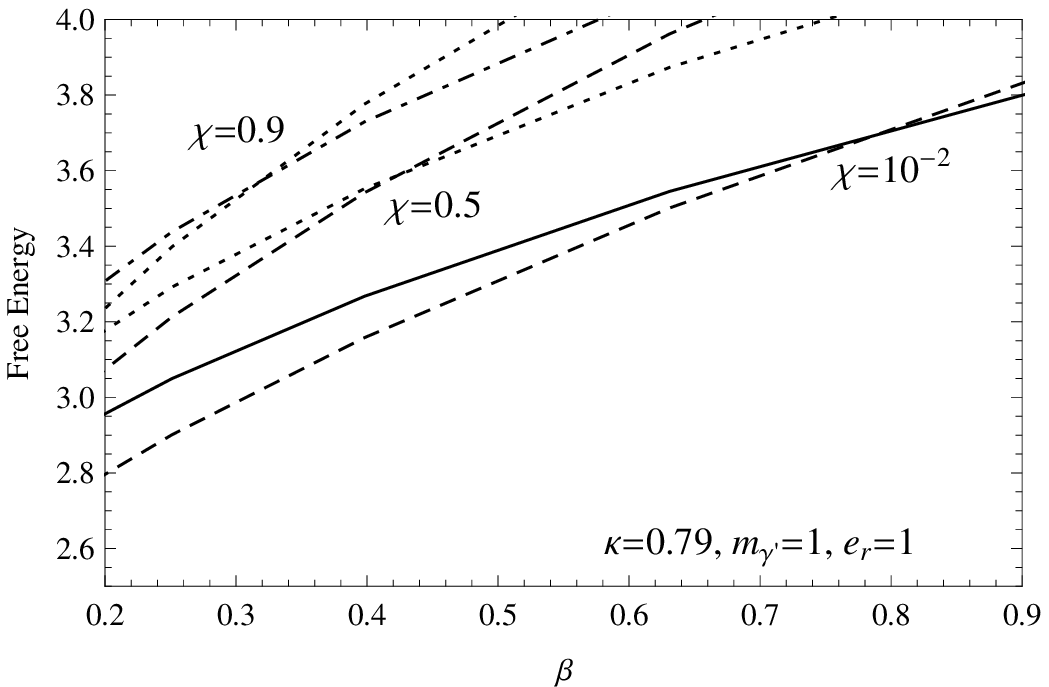}
\caption{Stability of vortices for different values of  the kinetic mixing parameter $\chi$. We have fixed $\kappa$ to the value corresponding
to the critical point for the case in which there is no mixing.
The rest of the parameters have been fixed to $\mu = e_r = 1$. Dashed lines correspond to the energy of a configuration $(2,2)$ while dotted ones correspond to twice the energy of configuration $(1,1)$. As $\chi$
increases, the critical point moves to the left.}
\label{fig:fig1}
\end{figure}

 For the case  $\chi n k >0 $, our results are shown in figure
  (\ref{fig:fig1}) where we plotted  the energy as a function of $\beta$ for  a (2,2)  vortex configuration  compared  to twice the energy of a (1,1) configuration  for different values of $\chi$.  {{Here our notation $(n,k)$ stands for the energy given by equation (\ref{energyvar}) with winding number $n$ of the visible sector and $k$ of the hidden one, respectively.}} We see that as $\chi$  grows the value of the critical point  beyond which the instability starts to move to lower and lower values of $\beta$.

\begin{figure}
\centering
\includegraphics[scale=0.7]{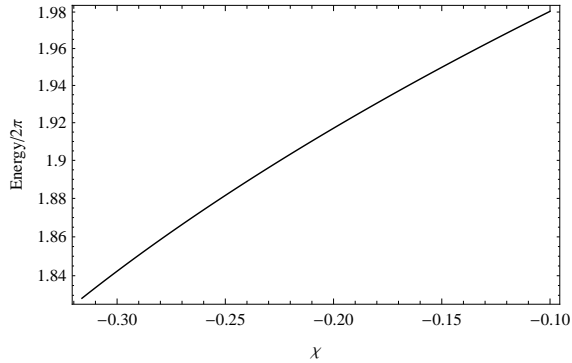}
     \caption{{Energy as a function of the mixing parameter  $\chi$ when $\chi<0$. The energy decreases as $\chi $ becomes more negative. The rest of the parameters have been fixed to unity, $\mu=e_r=\kappa=\beta$=1. }}
\label{fig:negative_chi}
\end{figure}

When $\chi < 0$ and $nk > 0$  the situation changes drastically. One can easily see  this  by considering  the particular {{limiting}} case  $\chi \rightarrow -1$, $nk >0$, and all other  physical parameters of the visible and hidden sector identical. With this choice, the gauge fields are indistinguishable and hence the first three terms in \eqref{redefenergy2} cancel out  and the energy will be smaller than {{the one}}   in which both signs coincide. {{Since}} there is  no contribution to the energy  from  the  visible and hidden field strengths, one should expect that the total energy  could become negative in some region of physical parameters  and vortex-like solutions will cease to exist. Our numerical analysis confirms  that this is indeed what happens, as can be seen in { figure~(\ref{fig:negative_chi})} where, as   $\chi$ approaches to -1 the energy becomes smaller,  until, in the region  $\chi \gtrsim -1$ it eventually becomes negative.

\begin{figure}
\centering
\includegraphics[scale=0.7]{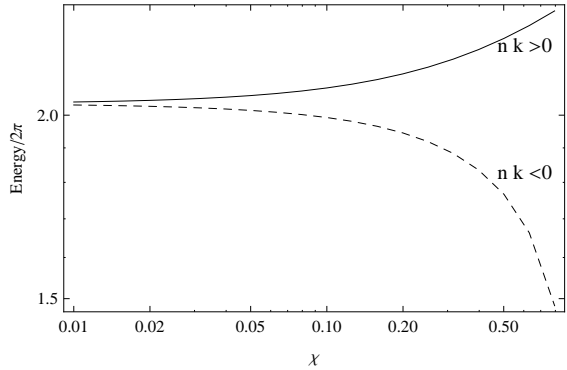}
     \caption{ {Energy vs. $\chi$ for a configuration with parallel fluxes, ${\rm sign}\; n= {\rm sign\; k}$,  and inverted fluxes, ${\rm sign}\; n \ne {\rm sign\; k}$. For higher values of $\chi$ the free energy has very different behavior for $nk>0$, where the energy grows with $\chi$, and $nk<0$, where the energy diminishes as $\chi$ grows. The parameters have been fixed to $|n|=|k|=1$,  and $e_r=\kappa=\beta=1$.}}
\label{fig:fig4}
\end{figure}

 If $\chi$ is still positive but $nk <0$, {\it{i.e}}  the magnetic fluxes from the hidden and visible sectors have opposite signs, the variational analysis shows in figure~(\ref{fig:fig4}) that when $nk<0$, the free energy diminishes as $\chi$ {{grows approaching one. This means that it is  favorable - when the mixing parameter is not negligible - to form vortices of opposite magnetic fluxes }}

\subsection{Changing {{the ratio}} $e_h/e \equiv e_r$}

 When the gauge couplings from visible and hidden sector are different the conclusion   concerning the stability of vortices  is similar to that  {{in subsection}} (4.2). To see this let us  fix the visible gauge coupling to $e=1$ and vary the corresponding hidden one.

\begin{figure}[h]
\begin{minipage}[b]{0.5\linewidth}
\centering
\includegraphics[width=\textwidth]{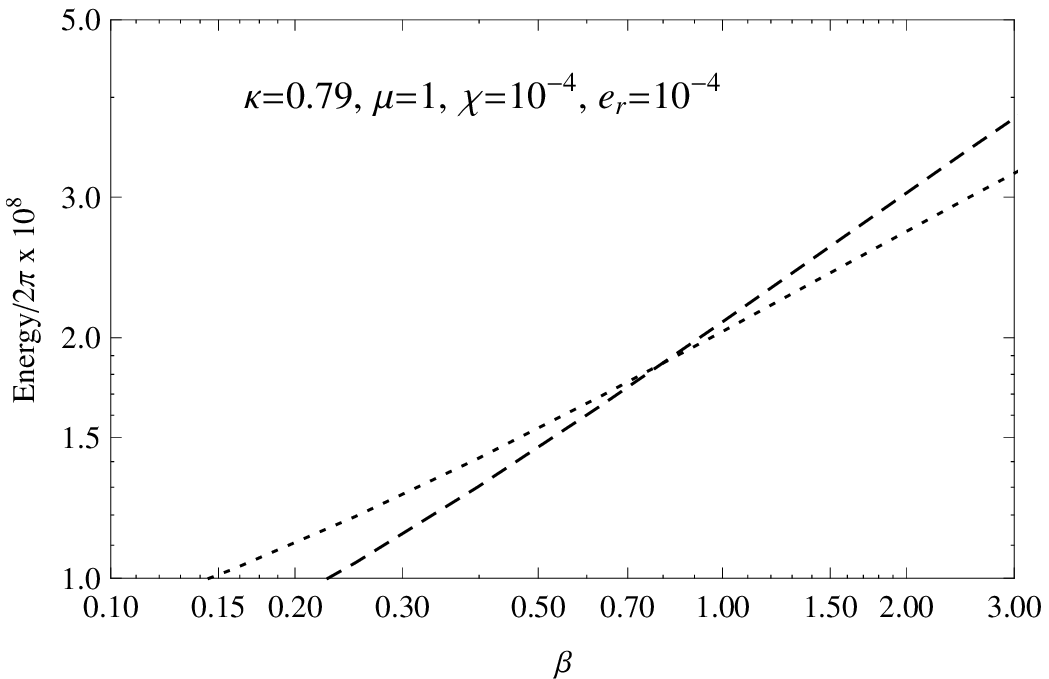}	
\caption{\footnotesize{For $e_r$ and $\chi$ very small
 the critical point does not change with respect to the case without mixing.  Dashed lines correspond to the energy of a configuration $(2,2)$ while dotted ones correspond to twice the energy of configuration $(1,1)$.}}
	\label{fig:fig2}
\end{minipage}
	\hspace{0.5cm}
\begin{minipage}[b]{0.5\linewidth}
\centering
\includegraphics[width=\textwidth]{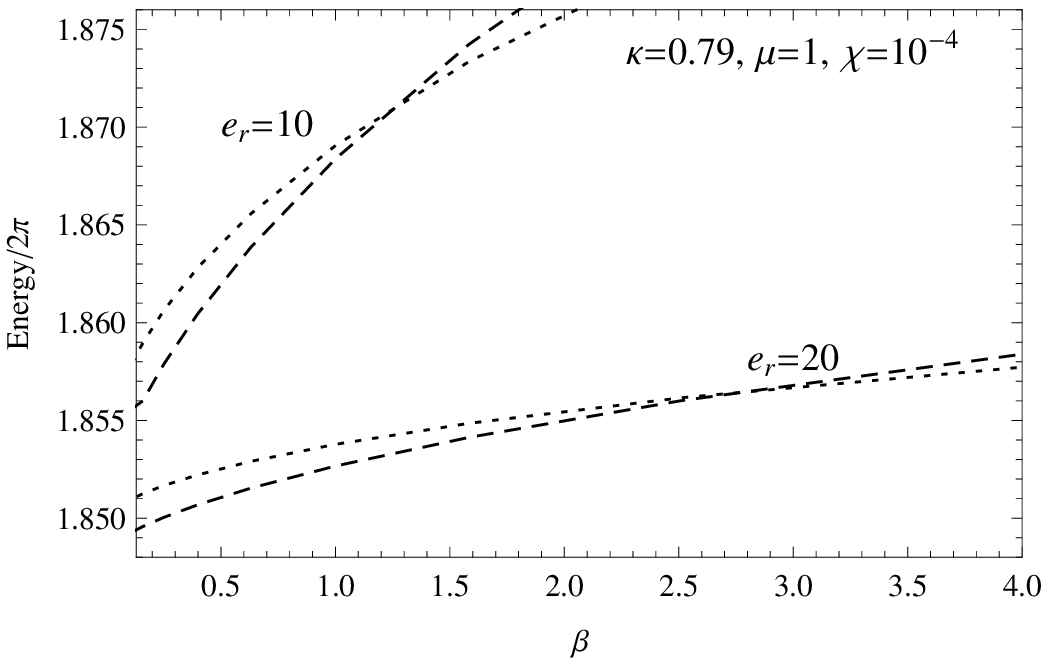}
	\caption{\footnotesize{For large $e_r$ and $\chi$ very small
 the critical point moves to the right with respect to the case without mixing. {Dashed lines correspond to the energy of a configuration $(2,2)$ while dotted ones correspond to twice the energy of configuration $(1,1)$.}}}
\label{fig:fig3}
\end{minipage}
\end{figure}

\begin{figure}[h]
\begin{minipage}[b]{0.5\linewidth}
\begin{center}
\includegraphics[width=\textwidth]{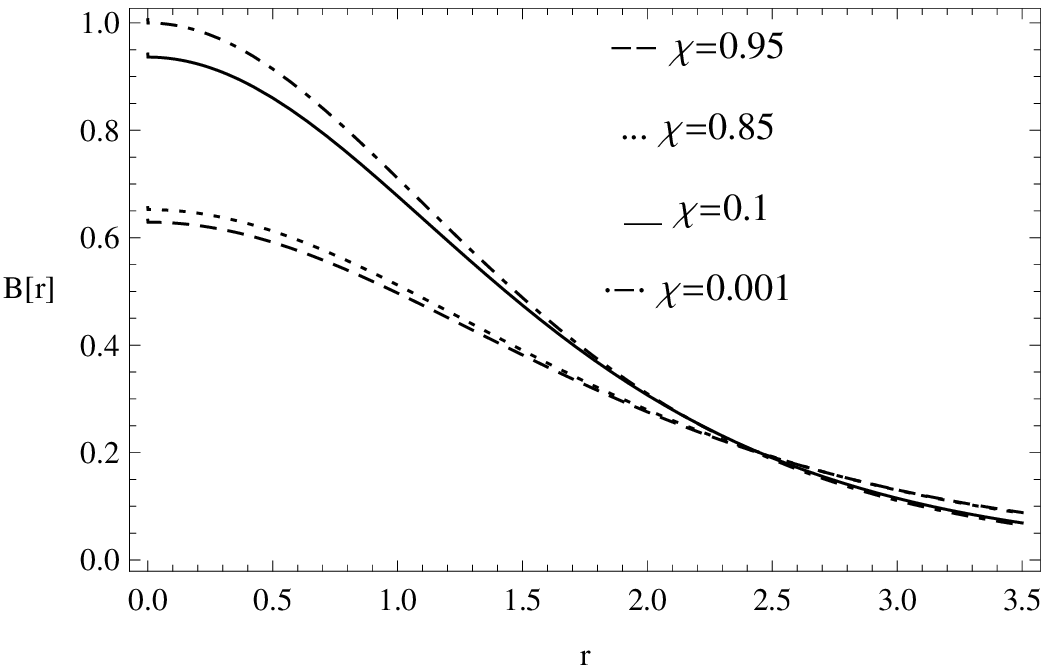}	
 {\footnotesize{(a)}}
\end{center}
\end{minipage}
	\hspace{0.3cm}
\begin{minipage}[b]{0.5\linewidth}
\begin{center}
\includegraphics[width=\textwidth]{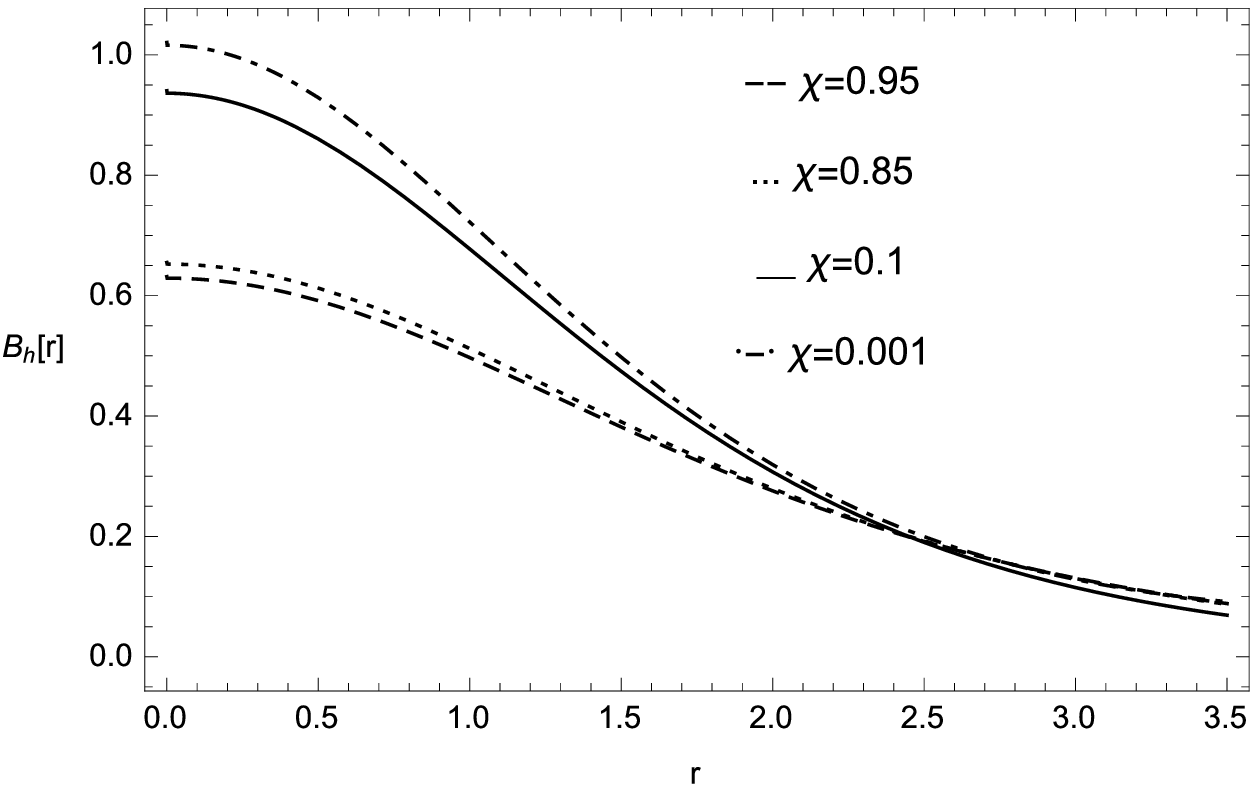}
{\footnotesize{(b)}}
\end{center}
\end{minipage}
	\caption{\footnotesize{(a): Visible magnetic field profiles for several values of the kinetic mixing parameter $\chi$.
	(b): Hidden magnetic field as a function of the kinetic mixing parameter $\chi$. The rest of the parameters have been fixed as $e_r=\mu=\kappa=\beta=1$, and the winding numbers $n=k=1$.	 }}
\label{fig:fig6}
\end{figure}

\begin{figure}[h]
\begin{minipage}[b]{0.5\linewidth}
\begin{center}
\includegraphics[width=\textwidth]{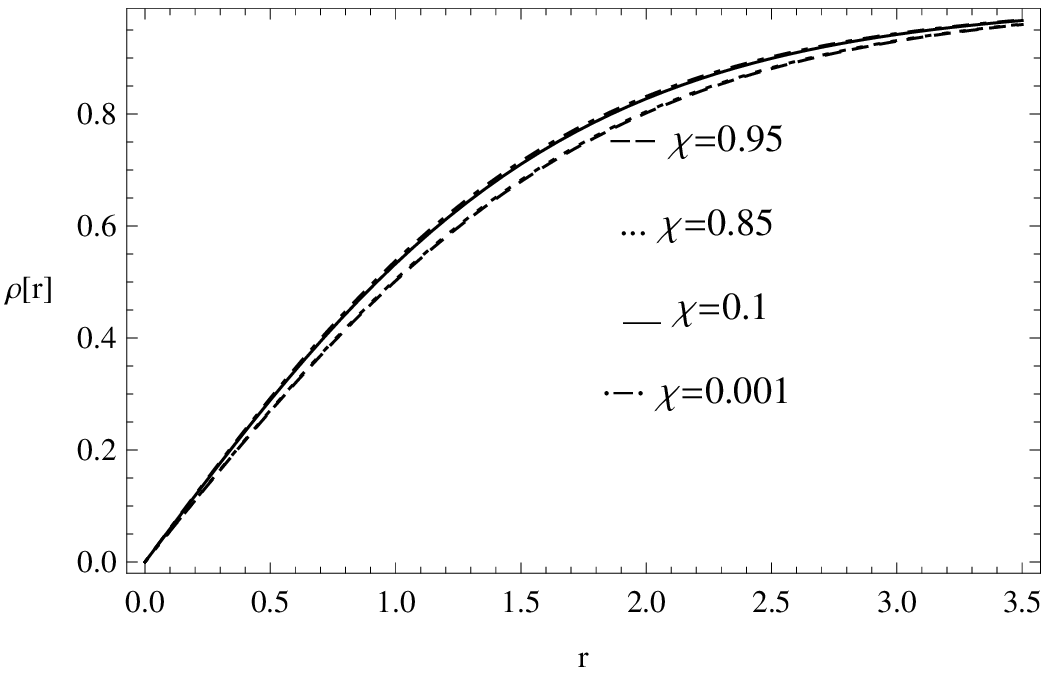}	
{\footnotesize{(a)}}
\end{center}
\end{minipage}
	\hspace{0.5cm}
\begin{minipage}[b]{0.5\linewidth}
\begin{center}
\includegraphics[width=\textwidth]{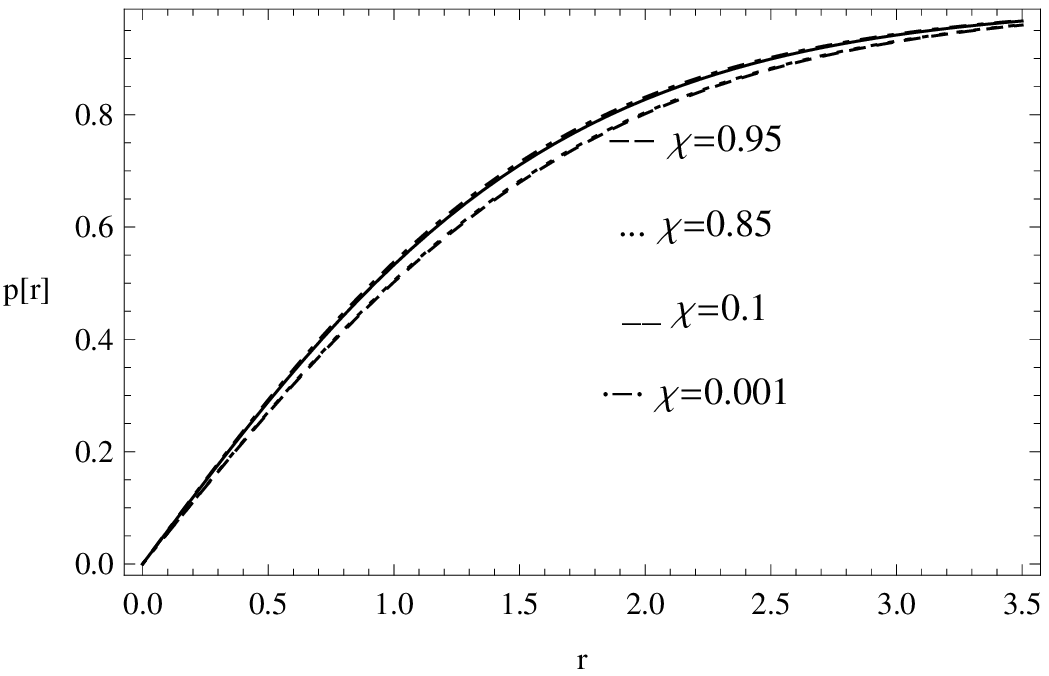}
{\footnotesize{(b)}}
\end{center}
\end{minipage}
	\caption{\footnotesize{(a): Visible Higgs field profiles for several values of the kinetic mixing parameter $\chi$.
	(b): Hidden Higgs field as a function of the kinetic mixing parameter $\chi$. The rest of the parameters have been fixed as $e_r=\mu=\kappa=\beta=1$, and the winding numbers $n=k=1$.	}}
\label{fig:higgs}
\end{figure}

We again chose to study the energy of a (2,2) vortex and compare it with twice the energy of a (1,1) vortex.
We   show  in figures (\ref{fig:fig2})-(\ref{fig:fig3}) the energy as a function of $\beta$ for $e_r>0$ (i.e. when both coupling constants have the same sign). Figure (\ref{fig:fig2}) shows that  for $e_h$ and $\chi$ very small ($e_h = \chi = 10^{-4}$) the critical stability point does not change with respect to the case without mixing. In contrast, when $e_h$  grows beyond
 the value $e_h = 1$, the critical  point moves to the right, as can be seen     in Figure  (\ref{fig:fig3}) for  $e_h=10,20$. {{Thus, as it was to be expected, only for large hidden gauge coupling charges vortex  stability is significantly affected.}

In the case ${\rm sign \,}e \ne {\rm sign \,}e_h$ (e.g.  $e_r=-1$), interesting phenomena take place for a suitable choice of the remaining parameters. To see this  let us consider a $\mathcal {CP}$ transformation of one of the fields, say  $\tilde G_\mu\equiv  \mathcal{CP}( G_\mu)=-G_\mu $ and choose $  \tilde G_\mu= A_\mu$. Then,  it is possible to get a cancelation of the kinetic terms for both vector fields when the physical parameters are chosen to be $\chi=\mu=1$\footnote{Note the condition 
$|\chi|<1$, previously found from asymptotic consistence does not hold in the present case.}. One could think of the above situation as describing a mixing between a gauge field  from the visible sector and an anti-hidden gauge field from the hidden sector {(of course this requires a definition of hidden field's antiparticles).}

{Now, when the gauge field kinetic terms cancel out, the field equation for the visible gauge field {{(which is identical to the CP transformed hidden one),}} reduces to
\be
ie\phi^* (\partial^\mu-ie A^\mu)\phi = 0,
\label{x}
\ee
so that just using the scalar field ansatz \eqref{ansatz1} one has, from the angular equation
\be
 (\partial_\varphi-ie A_\varphi)\phi = i\rho(r)(n - eA_\varphi) = 0,
\label{xx}
\ee
leading to
\bb
 A_\varphi= \frac{n}{r} =  \tilde G_\varphi.
 \label{esta}
\ee}
The singularity at the origin of both fields shows that - in the case of study - there are {no regular gauge field solutions}. Note that this singular solution for the gauge fields has been obtained
without  any reference to the scalar fields radial solution $\rho(r)$, since the corresponding field equation is completely decoupled   from the gauge field and depends only on the symmetry breaking potential. The only remnant of the gauge-scalar field interaction is the winding number $n$ appearing in eq.\,\eqref{esta} because of the phase in the scalar field ansatz.

{If one inserts the solution \eqref{esta} in
field equation for the Higgs scalar,
\be
D_\mu D^\mu\phi = \frac{\delta V[\phi]}{\delta \phi^*},
\ee
one just gets
\be
D_r D^r\phi = \frac{\delta V[\phi]}{\delta   \phi^*},
\ee
or, since $A_r=0$
\be
\rho(r)'' + \frac1r \rho' - \frac{\kappa^2}2(\rho^2 - 1) \rho = 0.
\label{simp}
\ee
(The same result can be obtained by making $\alpha =1$ in eq.\,\eqref{diffeq3}).}

Comparing eq.\eqref{simp} with the one corresponding to global vortices (see for example   \cite{global, Vilenkin}),
one can see that the only difference between the two is that
since there is no gauge field in the global $U(1)$, its radial scalar field equation contains extra term proportional to $n^2$
 which in our model's equation is canceled precisely by the contribution of $A_\varphi$. Precisely, due to the presence of this $n^2$ term, the global vortex energy diverges logarithmically \cite{hill}.

To see whether there is any energy divergence in our case we insert ansatz (\ref{ansatfi}), and the value of $A_\varphi$ given in  (\ref{esta}) in the energy per unit of length given by (\ref{redefenergy}). We get (for $n=1$)
\bb
\frac{E}{\phi_0^2\ell}=2\pi \left(\frac{1}8+\frac{89 \kappa^2}{1152 \mu^2} \right).
\ee
Hence, for any value of the variation parameter $\mu$, the above expression is finite. Now the minimum value of the energy corresponds to $\mu \rightarrow \infty$, so that $\rho $ becomes trivial, $\rho=1$, and $\phi=\phi_0 e^{in\varphi}$, an ill-defined expression at the origin. Thus, the energy per unit of length vanishes, and no regular non-trivial vortex solution therefore exist. The same conclusion holds for arbitrary value $n$. This result could have been obtained  just using the ordinary Bogomolny equations and replacing $\alpha(r)=1$,  which forces $\rho=1$.

Another interesting result correspond  to the case  $e_r=-1$. {Indeed, choosing the ansatz's radial functions $\gamma(r)=\alpha(r)$ and $\mu=\chi \rightarrow 1$}, a cancellation of the kinetic terms for gauge fields $\gamma$ and $\alpha$ also takes place. Moreover,  once again  a singular solutions for the gauge fields exist
{but consistency requires in this case an inverted magnetic flux condition imposing $n = -k$.}

\subsection{Radial dependence of fields}
In order to discuss  radial fields  profiles and their dependence
on the free parameters of the theory we shall follow two different numerical approaches: namely the variational approach already discussed and a shooting method.

{We start by varying the kinetic mixing parameter $\chi$, setting the rest of the parameters to unity,  $e_r=\beta=\kappa=\mu=1$ and the winding numbers $k=n=1$ so that visible and hidden fields are indistinguishable.
}

{In fig.~(\ref{fig:fig6}-a) we plot the visible magnetic fields obtained using the shooting method as a function of $r$ for several values of the kinetic mixing parameter $\chi$. We can conclude that increasing $\chi$ makes the magnitude of the magnetic field to decrease, thus lowering the visible magnetic energy.

Fig.~(\ref{fig:fig6}-b) shows  the hidden magnetic field as a function of $r$ for several values of $\chi$, using the shooting solution.{ Since the visible and hidden fields are indistinguishable, we obtain the same profile as the visible field.}
{In fig.~(\ref{fig:higgs}) we compare   the visible and hidden scalar fields as a function of $r$ for several values of $\chi$. From this graph we conclude that  as the  kinetic mixing parameter increases, the field reduces its asymptotic value.}

 Further, we have studied the behavior of the solution under the change of the mass ratio parameter  $\mu$ which has phenomenological relevance.
Note that for fixed $m_A=1$, increasing $\mu$ is equivalent to make the vacuum  value of the hidden Higgs field larger than the visible one.

 We have again taken $e_r=\beta=\kappa = 1$ , and a small value for  $\chi=10^{-4}$. The results we report were obtained using the variational approach, since for large $\mu$ it is more appropriate than the shooting method.  We plot  in figure (\ref{fig:Bmu})  the visible magnetic fields as a function of $r$ for several values of $\mu$. The plot suggests that when $\mu\geq 10 $   the visible magnetic field changes, both in magnitude and penetration depth.  This interesting result shows that a shorter range of the hidden field enforces the shortening  of the   visible range showing that non linear terms of the slowly decaying field affects the  $\mu \lesssim 5$ range, where both the shooting and the variational  methods are both applicable  their results  coincide}  showing that the visible magnetic field has the same behavior as the one where the visible sector has no mixing with a hidden sector .

{We have  also studied the field behavior under changes in $\chi/e_r$. From the analysis of the previous sections one can see  that this ratio can be regarded as an effective kinetic mixing which we shall call   $\chi_{\rm eff}\equiv \chi/e_r$. In particular,  using $\chi_{\rm eff}$ instead of just $\chi$   allows   to consider  more realistic values of the latter.}

{
The profiles of the visible magnetic field for different values of $\chi_{\rm eff}$ are shown in fig.~(\ref{fig:chieff}). Keeping  $\chi$   fixed to $\chi=10^{-5}$, we considered  different values $e_r$. Our results show that for a small $\chi_{\rm eff}$,  ($e_r \gg \chi$) the magnetic field shows no  departure from the behavior corresponding to the absence  of kinetic gauge mixing mixing with a hidden sector. However, as $\chi_{\rm eff}$ grows, ({\it{i.e.}}  $e_r \lesssim \chi$) the  magnetic field decreases but it has  a slower decay as $r$ grows.  For the curves of fig.~(\ref{fig:chieff}) we have fixed the rest of the physical parameters to unity.} {{Note that a value of $\chi_{\rm eff} > 1$ can be achieved by choosing small values of the kinetic mixing, for instance $\chi=10^{-7}$ and $e_r=10^{-8}$.}}

\subsection{Vortex decay into elementary configurations}
{Vortices with winding numbers $(n,k)$ could be unstable and decay into lower energy configurations, when available,  as it is the case in
 the ordinary Abelian Higgs model \cite{JR}. Indeed, in the absence of the hidden sector, the energy density in the type-II superconductivity vortex regime ($\kappa>1$) is proportional to
 the winding number squared, say $n^2$. Thus, a vortex with winding number $n=2$, will decay into two vortices of winding number $n=1$.} {{ We already studied the stability of the vortices on general grounds in sections 4.2 and 4.3.}}

When the mixing with the hidden sector is considered, the energy is no longer proportional to the two available winding numbers, $n^2, k^2$, but will also depend on the contribution of the mixing term,  which is related to  winding number $n$ and $k$ through the field strengths  and also depends on the values of parameters $\chi$ and $e_r$. In fact, we have seen in section 3.2 that vortex decay   depends crucially  on the sign of $\chi$.

We shall consider two types of elementary vortex configurations: the $(1,0)$ one carrying just  one unit of visible magnetic flux and the $(0,1)$ carrying instead just one unit of hidden magnetic flux. Then, starting with an $(n,k)$ configuration we shall analyze
under which conditions such configuration could decay  into one with $n$ elementary vortices of type  (1,0) and $k$ elementary vortices vortices $(0,1)$.
Let   us consider for definiteness the unbroken symmetry case discussed in our previous section. Taking for instance $ \phi=0 , \kappa= 0$ one can construct a  $k(1,0)$ configuration with $k$ spatially superimposed  hidden vortices  of unit flux. Then,  a vortex configuration of the type  $n(1,0)+k(0,1)$ can be formed by considering this configuration {and one where where the role of visible and hidden fields is inverted and  $k$  is replaced by  $n$.

We illustrate the decay   from the $(n,k)$ configuration as described above in  table~(\ref{tab:tabla1}) by comparing the energies of the $(2,2)$ configuration with that of  the $2(0,1)+2(0,1)$ one, for  different values of $\chi$ and  Landau parameters,  $\kappa, \beta$. As we can see, for small  values ($\chi  \sim 10^{-6}$) the decay of the configuration $(2,2)$ into the elementary ones   takes place approximately at the critical value of the Landau parameters if the  mixing were absent, that is, for $\kappa=\beta \sim 0.8 $. Now, as the mixing parameter grows, the decay takes place at lower and lower values of the Landau parameters. For instance, for $\chi\geq 0.5$ the decay of the vortex $(2,2)$   already occurs at $\kappa=\beta=0.6$.\\

 \begin{table}[t]
 \begin{adjustwidth}{1cm}{1cm}
\begin{center}
\begin{tabular}{ r  r  r  r }
   &  $\kappa=\beta=0.6$ &    \,\,\,\,\, & \,\,\,\,\,\,\,\,  $\kappa=\beta=0.8$
\end{tabular}
\begin{tabular}{ | c | c | c ||| c | c |}
  \hline
  $\chi$ &  (2,2)  & $2(0,1)+2(1,0)$ & (2,2) & $2(0,1)+2(1,0)$ \\
  \hline
  \hline
  $10^{-6}$ & 3.2007 &  3.3100& 3.7194 & 3.7163 \\
  \hline
  $10^{-3}$ & 3.2016 & 3.3100 & 3.7207 & 3.7164   \\
  \hline
  $10^{-1}$ & 3.2806 & 3.3034 & 3.8380 & 3.7088 \\
  \hline
  0.5 &  3.5569 & 3.1277 & 4.2366 & 3.5060 \\
  \hline
    \end{tabular}
    \end{center}
    \caption{\footnotesize{Energy of the  $(2,2)$ configuration (second and forth columns) and that of the $2(0,1) + 2(1,0)$ (third and fifth columns) for different values of the kinetic mixing parameter $\chi$, and two different values of the Landau parameters $\kappa=\beta$. The rest of the physical parameters have been fixed to $e_r=\mu=1$.}}
    \label{tab:tabla1}
    \end{adjustwidth}
\end{table}

Let us note that one can reach the same conclusion by varying  $e_r$  while keeping the kinetic mixing small, as discussed when we studied the radial fields profiles in terms of the effective mixing parameter $\chi_{\rm eff}$.  Note that for  phenomenologically  acceptable very small kinetic mixing parameter ($\chi \sim 10^{-6}$ or lower), the effect described above takes place when the hidden gauge coupling constant is very small,
$e_h/e \lesssim 10^{-6}$.

Finally, we have {investigated the effect of changes of the vector fields masses in the decay scenario. In table~(\ref{tab:tabla_mass})  the energies of a $(2,2)$ with a $2(0,1)+2(0,1)$ configuration are compared for several values of the mass ratio $\mu$ and for two different points of $(\kappa, \beta)$. One can see that increasing $\mu$ does not affect the stability of the vortices.}

 \begin{table}[t]
 \begin{adjustwidth}{1cm}{1cm}
\begin{center}
\begin{tabular}{ r  r  r  r }
   &  $\kappa=\beta=0.77$ &    \,\,\,\,\, & \,\,\,\,\,\,\,\,  $\kappa=\beta=0.8$
\end{tabular}
\begin{tabular}{ | c | c | c ||| c | c |}
  \hline
  $\mu$ &  (2,2)  & $2(0,1)+2(1,0)$ & (2,2) & $2(0,1)+2(1,0)$ \\
  \hline
  \hline
  $10^{-3}$ &1.82266 &1.85818 &1.85975 &1.85818  \\
  \hline
  $0.1$ &1.84087&  1.87647& 1.87832& 1.87676   \\
  \hline
  $1.0$ & 3.64538 & 3.68768 & 3.71954& 3.71635 \\
  \hline
  20 & 730.8801 &733.6594 & 750.5002 & 745.1293 \\
  \hline
    \end{tabular}
    \end{center}
    \caption{\footnotesize{Energy of the  $(2,2)$ configuration (second and fourth columns) and that of the $2(0,1) + 2(1,0)$ (third and fifth columns) for different values of the ratio of vector field masses, $\mu$, and two different values of the Landau parameters $\kappa=\beta=\left\{  0.77, 0.8   \right\}$. The rest of the physical parameters have been fixed to $e_r=1$ and $\chi=10^{-4}$.}}
    \label{tab:tabla_mass}
    \end{adjustwidth}
\end{table}

\section{The fields behavior in connection with superconductivity}

In view of the connection between the Landau-Ginzburg phenomenological theory for superconductors \cite{Ginzburg:1950sr} and the Abelian Higgs Model, superconductivity is a possible arena to test whether the mixing between the hidden and visible sectors could have a phenomenological impact. In this section we intend to give a brief and qualitative discussion on this issue.

{If one looks for measurable quantities that may have been affected  by the gauge mixing in the superconductivity context,  the scale lengths in the theory are natural the candidates to analyze.
In ordinary superconductivity (i.e., in the absence of a hidden sector) there are two characteristic lengths. One of them  is the penetration depth of the external magnetic field, $\ell$. In the language we have been using, it is given by the inverse of the {effective mass} of the gauge field, thus $\ell= m_A^{-1}$. The other one is the  characteristic length for the Cooper pairs, known as the {\it{coherence length}}, $\xi$ which in our notation would be $\xi=m_{\varphi}^{-1}$. These two lengths can be combined into one via the Landau parameter,  defined { {in our model}} as $\kappa=\ell^2/\xi^2$.
Thus, within a phenomenological   Ginzburg-Landau approach, there is only one free parameter, the Landau parameter which, after redefinitions of section 2,   { {is given by}} $\kappa=\sqrt{2\lambda}/{e}$.

{ The results obtained in  subsection {{4.4}} imply that when $\chi$ (or  $\chi_{\rm eff}$) approaches unity the visible fields get {{greatly}} modified, as it happens for large values of the  gauge boson masses ratio of $\mu$.
This means that depending on the  values of  physical parameters   ($\mu, e_r, \chi, \kappa, \beta$) the energy of a superconductor can get modified, thus  affecting the superconducting sample behavior, in particular  the exclusion of the magnetic field from it.}

{In order to analyze this issue we shall study the energy density behavior as a function of $r$  in the context of   superconductivity, when a mixing of visible photons with massive hidden photons through the kinetic mixing is present. We shall assume  for simplicity that energy density in  the superconductor  sample is governed - within the Ginzburg-Landau approach - by a the usual free energy density,  just composed of the visible magnetic field, the kinetic energy of the super current and the condensation energy of the Cooper pairs. The existence of a hidden sector will be taken into account by inserting in such free energy  the solutions  obtained by the minimization of the complete visible-hidden  model, eq.~(\ref{energy1}).
Then, the free energy density in the superconductor is taken as}
\bb
 \mathcal F^{visible}_{s}= \frac{B^2}2+\frac{1}2 |\partial_i \phi -i A_i \phi |^2+\frac{\kappa^2}8\left(|\phi|^2-1\right)^2,
\label{free_energy}
\ee
Note that with our conventions the {\it Landau parameter} is just $\kappa$ and the  Bogomolny point is $\kappa=1$.

We show in figure~(\ref{fig:energy_mass})  the energy density, (\ref{free_energy}) as a function of $r$ for several values of the hidden {{vector field}} mass.{ The continuous solid line in the figure corresponds to the case of an ordinary superconductor (i.e. in the absence of a hidden sector)}. As we can see, when the parameter $\mu$ is small ($\mu\lesssim 15$) there is no appreciable change {of the free energy} compared to the one where there is no mixing with a hidden sector. As}} this parameter grows, we observe a departure from the ordinary superconductor curve.  This result agrees with those reported in section {{4.}} For high values of $\mu$ the visible magnetic field increases its amplitude, thus increasing the magnetic energy, but   its penetration depth decreases. A similar conclusion should be reached by considering the energy density for different values of $\chi_{\rm eff}$.

\begin{table}[t]
\begin{adjustwidth}{1cm}{1cm}
\begin{center}
\begin{tabular}{ | c | c | }
  \hline
  $\chi$ &  $\sigma/2\pi$  \\
  \hline
  \hline
  $10^{-3}$ & 0.000003  \\
  \hline
  $10^{-2}$ & 0.000036 \\
  \hline
  $10^{-1}$ & 0.000709 \\
  \hline
  0.85 &  0.003882  \\
  \hline
  0.95 & 0.04066 \\
  \hline
    \end{tabular}
    \end{center}
    \caption{Surface energy for different values of the kinetic mixing parameter. As $\chi $ increases, the surface energy also increases. The remaining parameters have been fixed as $\kappa=\beta=e_r=\mu=1$.}
    \label{tab:tabla2}
    \end{adjustwidth}
\end{table}

The surface energy between a normal   and  superconducting samples is a relevant quantity in superconductivity since its sign unequivocally defines  the transition between type-I and type-II superconductivity.  The minimum of the surface energy occurs at the point where the free energy gets its minimum (where the Bogomolny bound is saturated), which in a normal Nielsen-Olesen vortex, with dimensionless variables, is $\kappa=1$.

{We have numerically studied the two dimensional surface  energy $\sigma$  associated to the visible sector of our model,  given by
\bb
\sigma=2\pi  \int_0^\infty \left(\frac{1}2 \left(B(r)-\frac{\kappa}{2}\right)^2-\frac{\kappa^2}8 \rho(r)^4\right)r dr
\label{dddd}
\ee
We see from this equation that $\sigma$   vanishes when $B(r)=\kappa/2 \left(1+\rho(r)^2\right)$, which is indeed the Bogomolny equation for the ordinary Abelian Higgs model  holding
 when $\kappa = 1$.

As stated above, the visible magnetic and scalars fields in eq.\eqref{dddd} correspond to the solutions of the complete set \eqref{diffeq1}-\eqref{disorder field}  that we found using an improved shooting method  in order to  refine  accuracy of the calculation. To determine how the surface energy measured in experiment could be affected by the existence of a hidden sector, we have varied the free parameters of our model and made use of the equation for the surface energy.}} We show in table~(\ref{tab:tabla2}) the value of the surface energy when $\kappa=1$ for different values of the parameter $\chi$. The rest of the phenomenological parameters were fixed to $\beta=e_r=\mu=1$. We clearly see that increasing the value of $\chi$ makes the surface energy at $\kappa=1$ to grow.  This result can be interpreted as a shift in the value of the limiting point between type-I and type-II superconductivity supporting   our previous statement on the non-existence of first order Bogomolny equations.

\begin{table}[t]
\begin{adjustwidth}{1cm}{1cm}
\begin{center}
\begin{tabular}{ | c | c | }
  \hline
  $\chi_{\rm eff}$ &  $\sigma/2\pi$  \\
  \hline
  \hline
  $0.01$ & 0.000242  \\
  \hline
  $0.1$ & 0.007925 \\
  \hline
  $0.9$ & 0.058660 \\
  \hline
  1.0 &  0.069072 \\
  \hline
  1.25 & 0.102120 \\
  \hline
    \end{tabular}
    \end{center}
    \caption{Surface energy for different values of the effective kinetic mixing parameter. As $\chi_{\rm eft}$ increases, the surface energy also increases. For all the values shown here we have considered a kinetic mixing of $\chi=10^{-5}$. The remaining parameters have been fixed as $\kappa=\beta=\mu=1$.}
    \label{tab:tabla3}
     \end{adjustwidth}
\end{table}

From the experimental point of view such large values of the kinetic mixing parameter $\chi$
 have been ruled out by experiments \cite{Essig:2013lka}. In view of this, we have computed the surface energy in terms of the {\it{effective}} kinetic mixing  $\chi_{\rm eff}\equiv \chi/e_r$. In this way  we can consider more realistic values $\chi$ taking a small value for the  hidden gauge coupling compared to the visible one. In Table~(\ref{tab:tabla3}) we show the surface energy, eq.~(\ref{dddd}) for different values of the effective kinetic mixing. One can see that even for small kinetic mixing   the surface energy now changes appreciably

Concerning the visible magnetic field profiles, the results plotted in figure~(\ref{fig:Bmu}) suggest that the point at which  the surface energy vanishes also changes as $\mu$ grows
significantly

\section{Summary and discussion}
In this work we have analyzed a  gauge theory with a visible and a hidden sector with dynamics governed by two Abelian Higgs Lagrangians coupled through a gauge kinetic mixing. Imposing the usual cylindrically symmetric Nielsen-Olesen ansatz for gauge and scalar fields  we have arrived to a system of 4 coupled radial equations which we have solved numerically.

 We have started studying the case in which the $U(1)$ gauge symmetry is unbroken in one of the two sectors. This was achieved    by not including  in the Lagrangian the corresponding  complex scalar. We found  that even in this case  the kinetic gauge mixing forces the existence of vortex configurations also in  the unbroken sector  with associated  magnetic field decaying exponentially at infinity with the  same length as the one in the broken symmetry sector.

Interestingly, always in the case in which one U(1)  symmetry is unbroken, gauge and scalar self-interaction  coupling constants satisfy a relation which depends on the value of gauge mixing parameter $\chi$ and  first order  Bogomolny equations    exist in the  broken sector. The fact that the two fields strengths are proportional (with a proportionality factor    $(e\chi/e_h)$) explains why both magnetic fields have the same exponential decay.
This is a relevant result that could be in principle exploited considering  for instance primordial magnetic fields generation by  dark superconducting strings in the early universe \cite{Vachaspati:1991nm}.

Concerning the case in which both $U(1)$ gauge symmetries are broken,     we have found that the relevant parameters controlling stability are
  $\chi n k$ (with $n$ and $k$ the units of magnetic flux) and the ratio of the gauge couplings $e_r = e_h/e$. Our numerical analysis shows that
for growing values of $\chi n k >0$  and  $e_r>0$, the instability regime starts at lower values of
the hidden sector of the Landau parameter.
 If $\chi$ is instead positive but $nk <0$ with  $e_r$ positive  we find that the energy gets reduced as the parameter $\chi$ grows, the opposite of the $\chi n k>0$ case.

 We also  studied  the dependence of the solutions on the gauge coupling constants ratio $e_r$. To this end we considered the case of small $\chi \sim 10^{-4}$ so as to detect the individual dependence on $e_r$. When $e_r>0$,  for  very small values of $e_r$  ($e_r =  10^{-4}$)  the critical stability point does
not   change significantly compared to the case with no mixing. In contrast, when  $e_r > 1 $   the decay critical stability point   moves to the right  as  $e_r$ grows.

{
Interesting phenomena take place when ${\rm sign \,}e \ne {\rm sign \,}e_h$ (i.e. $e_r<0$) together with suitable  choices of the remaining parameters. In particular if  the $\mathcal{CP}$ transformed  hidden gauge field   is equal to the visible one, $\mathcal{CP}( G_\mu)=A_\mu$,   the kinetic terms for both vector fields  cancel out for  $\chi \rightarrow 1$  and $m_A = m_G$. This identification can be interpreted in terms of a    mixing between a photon from the visible sector and an anti-hidden photon from the hidden sector (of course this requires a definition of hidden field's antiparticles). Being the gauge  kinetic terms absent, one finds a solution of the form   $\phi = \phi_0\exp(in\varphi)$ and $A_\varphi = n/r$. That is, both fields are singular at the origin but the singularities cancel out when computing the energy per unit length.}

We have found that both hidden and visible  magnetic fields reduce their magnitude when $\chi$ or $\chi_{\rm eff}$ approaches   unity.  In respect to
  changes in $\mu$,  the variational method shows  observable effects in the  visible magnetic fields when $\mu \gtrsim 15$. Concerning the    hidden  magnetic field, it  grows significantly for   $\mu \gtrsim 1$.

Concerning the decay of  $(n,k)$  vortices     we have studied the case in which the final configuration is a combination of $n (1,0)$ and $k  (0,1)$ elementary vortices. The conclusion is that  as the gauge mixing parameter grows, the decay takes place at lower and lower values of the hidden Landau parameter with the visible one is fixed. The same holds if one varies
  $\chi_{\rm eff}$ or $1/e_r$. Using a phenomenologically  acceptable kinetic mixing parameter ($\chi \sim 10^{-6}$)  the effect described above takes place when the hidden gauge coupling constant satisfies $e_h/e \lesssim 10^{-6}$.

We have also presented a qualitative discussion of the results from previous sections in connection with superconductivity.  As expected, for small $\chi$ the results remain  unchanged with respect  to the case in which no hidden sector is present. We have shown that  the mass ratio $\mu$ and  effective gauge kinetic mixing $\chi_{\rm eff}$ are the relevant parameters to study the hidden sector effect on a superconductor sample. Concerning the former,  we   found  that the energy density grows when $\mu$ increases, but the effective penetration length   is reduced.
 In the normal superconducting theory the surface energy is zero at the Bogomolny point $\kappa=1$. However, in the presence of a gauge mixing, when $\chi$ or $\chi_{\rm eff}$ approach unity the surface energy changes its behavior and  does not vanish for  $\kappa=1$.

We conclude that in view of the very rich structure of the vortex solution space that we have found, it would be worthwhile to analyze the  role of the vortex configurations in cosmology, hidden photon search and supersymmetric extensions. We expect to discuss these issues in a future work.

\section*{Acknowledgements}
P.A. was supported by FONDECYT project 11121403 and Anillo ACT 1102. F.A.S is financially supported by CONICET, ANPCIT, UNLP and CICBA grants

{We are  specially thankful of  E. Moreno for his useful comments and help in the numerical calculations, J.~Jaeckel, for reading the manuscript and for his valuable comments,  J.~Gamboa for his suggestion and encouragement to look into this subject, and A. Ringwald and J. Redondo for their participation in earlier stages of this work. We are also thankful of G. D{u}ering, G. Lozano and  E. Mu\~noz for  discussions and comments.}
\newpage
\section*{Appendix: Asymptotic behavior of the radial fields}
We find numerical solutions of the radial equations eqs.~(\ref{diffeq1})-(\ref{disorder field}) by implementing a shooting method to match the solutions of these equations in the limit $r\rightarrow \infty$. In order to find the analytical asymptotic solutions of these equations, we start by defining the functions  $\tilde \alpha= \alpha-1$ and $\tilde \gamma= \gamma-1$, $\tilde \rho=\rho-1$, $\tilde p=p-\frac{\mu}{e_r}$ such that in the limit $r\rightarrow \infty$, they all satisfy
\begin{eqnarray}
 \lim_{r \to \infty} \tilde \rho(r) = 0 \;, \;\;\; \lim_{r \to \infty}  \tilde p(r) =0. \nonumber\\
 \lim_{r \to \infty} \tilde \alpha(r)  =0  \;, \;\;\;  \lim_{r \to \infty}  \tilde \gamma(r) = 0.
 \label{boundary2}
\end{eqnarray}
With these redefinitions, eqs.~(\ref{diffeq1})-(\ref{disorder field}){ take in the asymptotic limit the form}
\eqb
\left[ r\frac{d}{dr}  \left(\frac{1}r \frac{d}{dr}\right) \right] \left(n \tilde \alpha+\frac{k}{e_r}\chi \tilde \gamma \right)  -n \tilde \alpha &=& 0,
\label{44}\\
\left[ r\frac{d}{dr}  \left(\frac{1}r \frac{d}{dr}\right) \right] \left( \frac{k}{e_r} \tilde \gamma +n \chi \tilde \alpha \right)  - \frac{k}{e_r}\mu^2 \tilde \gamma &=& 0, \label{45}\\
\tilde \rho''+\frac{\tilde \rho}{r}-\kappa^2\tilde \rho&=&0,\\
\tilde p''+\frac{\tilde p}{r}-\left(\beta \mu\right)^2\tilde p&=&0.\\
\eqf
{The solutions for $\tilde \rho$ and $\tilde p$ are}
 \eqb
 \tilde \rho(r)&=& D_1 K_0 (\kappa r)+D_2 I_0(\kappa r),\\
 \tilde p(r)&=& E_1 K_0(\mu \beta r)+ E_2 I_0(\mu \beta r).
 \eqf
{Making $n \tilde \alpha \to \tilde \alpha $   and ${k}{/e_r} \tilde \gamma \to \tilde \gamma$   eqs.\eqref{44}-\eqref{45} become}
 \eqb
\left[ r\frac{d}{dr}  \left(\frac{1}r \frac{d}{dr}\right) \right] \left( \tilde \alpha+\chi \tilde \gamma \right)  - \tilde \alpha &=& 0,\\
\left[ r\frac{d}{dr}  \left(\frac{1}r \frac{d}{dr}\right) \right] \left(  \tilde \gamma +\chi \tilde \alpha \right)  -\mu^2 \tilde \gamma &=& 0,
\eqf
which can be combined into the equation
\eqb
\left[ r\frac{d}{dr}  \left(\frac{1}r \frac{d}{dr}\right) \right] \left( \tilde \alpha \left(A+\chi B \right) + \tilde \gamma \left(B+ \chi A \right) \right)  = A \tilde \alpha +B \mu^2 \tilde \gamma, \label{desacoplo}
\eqf
{where $A, B$ are arbitrary constants.
We now introduce  $C$}
\eqb
A+\chi B&=& CA,\\
B+\chi A&=& CB\mu^2,
\eqf
{and solve for $A$ finding}
\bb
\frac{A_\pm}B= \frac{\mu^2-1}{2\chi}\pm \frac{1}{2\chi} \sqrt{\left(1-\mu^2\right)^2+4\mu^2 \chi^2}.
\ee
{So that $C_\pm$ can be written as}
\bb
C_{\pm}= \frac{1}{2\mu^2}\left(\mu^2+1\pm \sqrt{\left(1-\mu^2\right)^2+4\mu^2 \chi^2}\right). \label{ce}
\ee
With this  eq.(\ref{desacoplo}) becomes (for $\chi \neq 1$)
\bb
\left[ r\frac{d}{dr}  \left(\frac{1}r \frac{d}{dr}\right) \right]  F_{\pm}(r)= \frac{1}{\sqrt{C_\pm}}F_{\pm},
\label{bessels}
\ee
where the functions $F_\pm(r)$ are defined as
\bb
F_{\pm} (r)= \frac{A_\pm} B \tilde \alpha +\mu^2 \tilde \gamma. \label{ffunction}
\ee
{The solution of equation (\ref{bessels}) is then}
\eqb
F_+ (r) &=& A_1 r K_1\left(\frac{r}{\sqrt{C_+}}\right) +A_2 r I_1\left(\frac{r}{\sqrt{C_+}}\right), \\
F_- (r) &=& B_1 r K_1\left(\frac{r}{\sqrt{C_-}}\right)+B_2 r I_1\left(\frac{r}{\sqrt{C_-}}\right).
\eqf
Form this result one gets  $\tilde \alpha$ and $\tilde \gamma$ in the asymptotic limit $r\rightarrow \infty$
\eqb
\tilde \alpha&=& n\frac{F_+ - F_-}{A_+/B- A_-/B},\\
\tilde \gamma&=& \frac{k}{e_r} \frac{\left((A_-/B) F_+ - (A_+/B) F_-\right)}{\mu^2\left(A_+/B- A_-/B \right)}.
\eqf

{Now,  in order to have exponential decays for the massive fields at $r\rightarrow \infty$ one should impose $C_\pm >0$, which in turn implies
\bb
\left(1+\mu^2\right)^2 > \left(1-\mu^2\right)^2+4\mu^2 \chi^2,
\ee
or
\bb
 \chi^2 <1.  \label{condition}
\ee
 This is an important result showing that in order to have finite energy vortex solutions   parameter $\chi$ controlling the mixing between the visible and the hidden sectors should satisfy $|\chi| <1$.}

\newpage


\begin{figure}[h]
\begin{minipage}[b]{0.5\linewidth}
\begin{center}
\includegraphics[width=\textwidth]{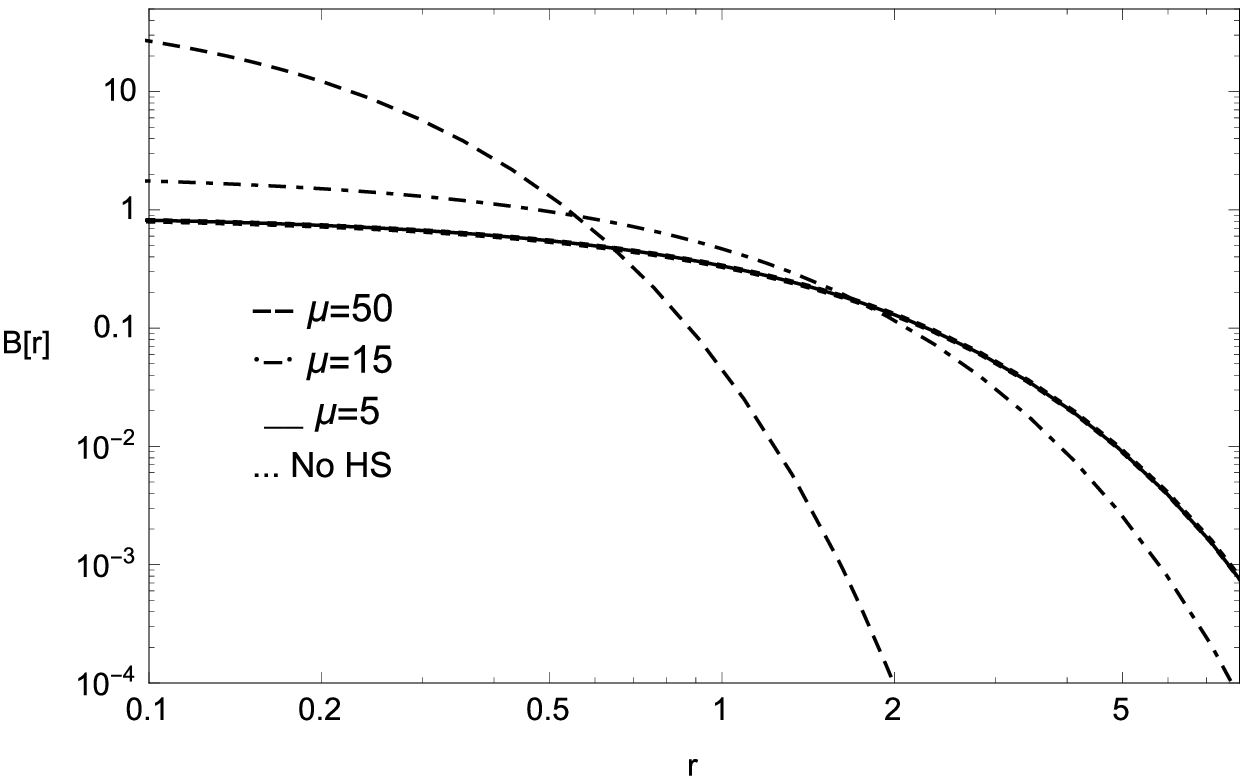}	
{\footnotesize{(a)}}
\end{center}
\end{minipage}
	\hspace{0.5cm}
\begin{minipage}[b]{0.5\linewidth}
\begin{center}
\includegraphics[width=\textwidth]{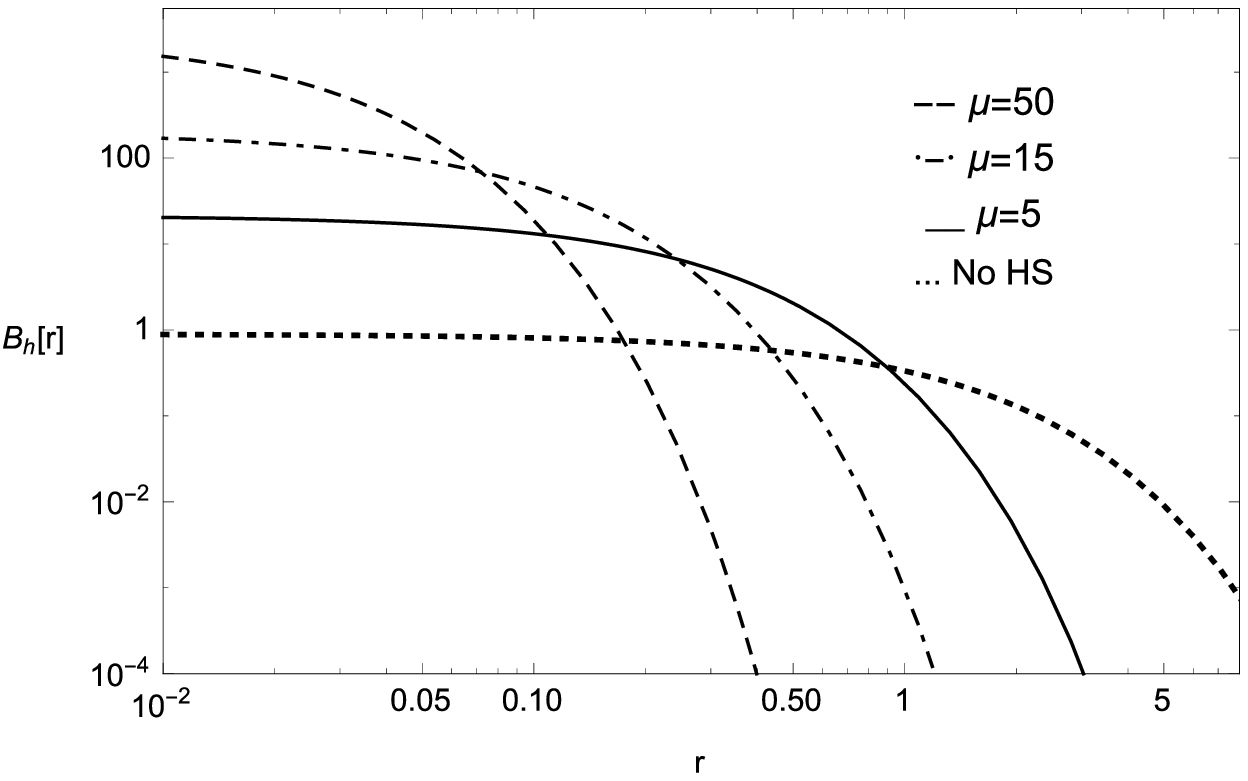}
{\footnotesize{(b)}}
\end{center}
\end{minipage}
	\caption{\footnotesize{(a): Visible magnetic field profiles for several gauge bosons masses, $\mu$.
	(b): Hidden magnetic field profiles for several values of  $\mu$. Both plots have been obtained using the variational method. The rest of the parameters have been fixed as $e_r=\kappa=\beta=1$, $\chi=10^{-4}$, and the winding numbers $n=k=1$.	}}
\label{fig:Bmu}
\end{figure}

\begin{figure}
\centering
\includegraphics[scale=0.7]{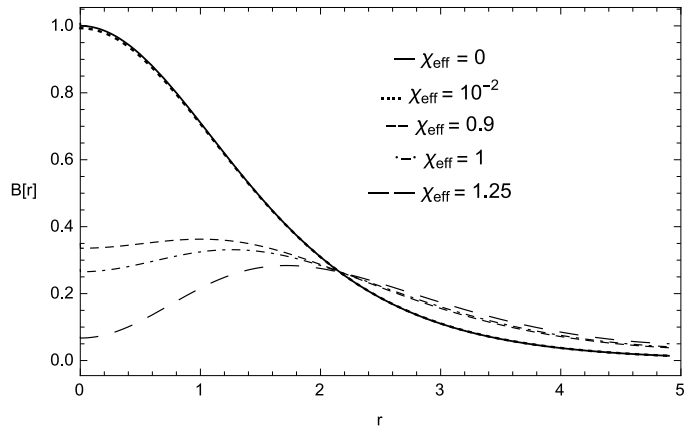}
     \caption{ {Visible magnetic field profile for different values of the effective kinetic mixing $\chi_{\rm eff}=\chi/e_r$. For all curves, the kinetic mixing is fixed to $\chi=10^{-5}$. The rest of the parameters remain as $\mu=\kappa=\beta=1$, with winding numbers $n=k=1$. }}
\label{fig:chieff}
\end{figure}

\begin{figure}[t]
\centering
\includegraphics[scale=0.7]{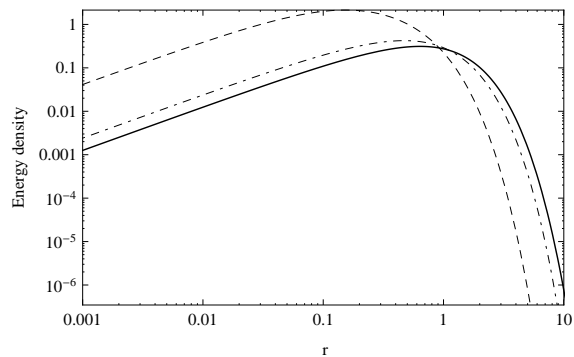}
     \caption{{Energy density profiles for several values of $\mu$ (we have set $\kappa=1$). The continuous solid line corresponds to the energy in the case of no mixing with a hidden sector.  The dot-dashed line corresponds to $\mu=15$ and the dashed line corresponds to $\mu=20$. {{Thus, when the mass of the hidden gauge boson grows significantly from unity, the energy density of the superconductor departs from the value of the usual Ginzburg-Landau theory.}}  The rest of the parameters have been chosen as: $\beta=\kappa=e_r=1$, and $\chi=10^{-4}$.  }}
\label{fig:energy_mass}
\end{figure}

\end{document}